\documentclass[aps,pre,reprint, superscriptaddress]{revtex4-2}
\usepackage{blindtext}
\usepackage{graphicx}
\usepackage{amsmath}
\usepackage{amsthm}
\usepackage{amssymb}
\usepackage{amsfonts}
\usepackage{siunitx}

\begin{document}

\title{White-noise fluctuation theorem for Langevin dynamics}
\author{M. Innerbichler}
\affiliation{Faculty of Physics, University of Vienna, 1090 Vienna, Austria}
\author{A. Militaru}
\affiliation{Photonics Laboratory, ETH Z\"urich, CH-8093 Zurich, Switzerland}
\author{M. Frimmer}
\affiliation{Photonics Laboratory, ETH Z\"urich, CH-8093 Zurich, Switzerland}
\author{L. Novotny}
\affiliation{Photonics Laboratory, ETH Z\"urich, CH-8093 Zurich, Switzerland}
\affiliation{Quantum Center, ETH Z\"urich, CH-8093 Zurich, Switzerland}
\author{C. Dellago}
\thanks{e-mail: christoph.dellago@univie.ac.at}
\affiliation{Faculty of Physics, University of Vienna, 1090 Vienna, Austria}

\begin{abstract}
	Fluctuation theorems based on time-reversal have provided remarkable insight into the non-equilibrium statistics of thermodynamic quantities like heat, work, and entropy production. These types of laws impose constraints on the distributions of certain trajectory functionals that reflect underlying dynamical symmetries. In this work, we introduce a detailed fluctuation theorem for Langevin dynamics that follows from the statistics of Gaussian white noise rather than from time-reversal. The theorem, which originates from a point-wise symmetry in phase space, holds individually for each degree of freedom coupled to additive or multiplicative noise. The relation is independent of the phase space distribution generated by the dynamics and can be used to derive a versatile parameter inference algorithm applicable to the a wide range of systems, including non-conservative and non-Markovian ones.
\end{abstract}

\maketitle

\section{Introduction}

Stochastic thermodynamics revolves around the generalization of macroscopic thermodynamics to the micro- and mesoscopic regime. At this scale, quantities like heat, work, or entropy production become functionals of stochastically evolving trajectories and typical fluctuations exhibit magnitudes comparable to the respective mean and standard deviation \cite{seifert2019stochastic}. Hence, it is necessary to consider probability distributions rather than operating with average values only. Remarkably, the fundamental symmetries of the underlying dynamics often result in so-called fluctuation theorems (FTs), i.e.\ physical laws constraining these distributions \cite{seifert2012stochastic, harris2007fluctuation}. Time-reversal in particular has led to several important relations like the well-known Crooks fluctuation theorem \cite{crooks1999entropy}: under the assumption of microscopic reversibility (detailed balance), this theorem connects the probability distributions of work dissipated during a non-equilibrium transformation carried out forward and backward in time. Other FTs constrain only expectation values: the Jarzynski equality \cite{jarzynski1997nonequilibrium} relates the exponentiated work average to the free energy difference between equilibrium states and the Hatano-Sasa relation \cite{hatano2001steady} extends this result to transitions between non-equilibrium stationary states, to name but a few \cite{seifert2012stochastic}.

Although various notions of time-reversal have been exploited to deduce FTs \cite{harris2007fluctuation}, they are not the only dynamical symmetries that permit doing so. Here, we explore this possibility for the case of general Langevin equations, which are widely used to model continuous stochastic processes and most commonly feature Gaussian white noise as a source of randomness \cite{langevin1908theorie, vanKampen1981stochastic}. The model in this particular form and its equivalent Fokker-Planck representation \cite{risken1989fpe} have been applied to a wide range of physical, chemical, and biological phenomena. Examples include micro- and macroscopic systems under the influence of thermal noise \cite{coffey2004langevin}, active particles \cite{schimansky1995structure, romanczuk2012active, fodor2016far}, reaction coordinates in transition processes \cite{berezhkovskii2013diffusion, peters2013reaction}, and the concentrations of reactants in a well-stirred solution \cite{gillespie2000chemical}. 

The relations introduced in this work have the form of detailed fluctuation theorems (DFTs) and emerge from the properties of Gaussian white noise without any reference to time-reversal. Seifert has previously discussed such relations for pure Langevin systems with arbitrary deterministic forces and additive noise \cite{seifert2008stochastic, seifert2012stochastic}. 
Here, we extend upon these investigations in several ways, also examining systems that are only partially described by Langevin equations with additive or multiplicative noise. We do not require the existence of a stationary state or any knowledge about the phase space distribution at any time. These white-noise DFTs offer a considerable degree of flexibility and remain valid for various systems that have previously been studied in search for reversibility-based distributional symmetries and inequalities. Examples range from magnetic forces \cite{jayannavar2007charged, wang2014fluctuation, coretti2021fluctuation} to active matter \cite{mandal2017entropy, dabelow2019irreversibility} and even delayed feedback \cite{rosinberg2015stochastic, rosinberg2017stochastic}. The treatment of each of these cases poses unique challenges in the framework of stochastic thermodynamics due to the specific nature of time-reversal. Conversely, the noise-symmetry approach followed here circumvents these additional complications per construction, demanding no special attention for any of the issues mentioned above.

Note, however, that the generality of the white-noise DFT does come at a cost: lacking the usual time-reversal--based connection to entropy production, the theorem does not provide any immediate thermodynamic insight. Moreover, it does not even require the existence of an underlying Hamiltonian to drive the system's dynamics, and may consequently apply to non-physical systems as well. 
Rather than refining our understanding of thermodynamics in particular, we use our results to devise a versatile parameter inference method building on the white-noise DFT. The general idea of this approach is to determine unknown parameters by minimizing the deviation between the statistics of the empirical data and the symmetry dictated by the DFT. We illustrate the method numerically on the example of an oscillator with delayed feedback, extracting the delay time. The method is widely applicable and only requires accurate trajectory data.

The remainder of this paper is organized as follows. In Sec.~\ref{sec:DFT}, we first derive a local version of the DFT for displacements conditioned to individual points in phase space.
We then extend the local version of the DFT to trajectory data, deriving Eq.~\eqref{eq: general WNDFT} which is the main result of this work.
We then demonstrate the validity of the DFT for several numerical examples in Secs.~\ref{Sec:Overdamped_Systems} and \ref{Sec:Underdamped_Systems}. These systems feature arbitrary initial distributions, lack of stationary densities, and non-Markovianity. In Sec.~\ref{Sec:Inference} we outline the parameter inference method and discuss its application. After reviewing the impact of signal contamination and measurement noise on the white-noise DFT in Sec.~\ref{sec:signalContamination}, we provide our concluding remarks in Sec.~\ref{Sec:Conclusions}.

\section{Noise-induced Fluctuation Theorem Symmetries}
\label{sec:DFT}

Let us consider a general system described by $M$ degrees of freedom $\mathbf{q}=\{q_1,q_2,\dots q_M\}$. These coordinates $q_i$ are in principle not restricted to positions and velocities, but may include more general quantities as well, e.g.\ reaction coordinates. The evolution of $\mathbf{q}$ can generally be described by an independent system of $M$ coupled equations. We assume that the first $N\leq M$ degrees of freedom evolve according to Langevin equations, that is laws of the form
\begin{equation}
	\dot{q_i} = f_i(\mathbf{q},t)+\sum_{j=1}^k g_{ij}(\mathbf{q},t)\eta_j(t).
	\label{eq:langevin}
\end{equation}
Here, $\dot{q_i}$ designates the derivative with respect to time $t$ of coordinate $q_i$, the $i$-th element of state vector $\mathbf{q}$. The index $j$, on the other hand, refers to the Gaussian white noise $\eta_j$. The $k$ distinct Gaussian white noises are independent, satisfying the relation $\langle \eta_j(t)\eta_{j'}(t') \rangle = \delta_{jj'}\delta(t-t')$. Finally, the functions $f_i$ and $g_{ij}$ prescribe the deterministic drift and noise amplitudes respectively. We deal with multiplicative noise if any $g_{ij}$ explicitly depends on $\mathbf{q}$. Note that Eq.~\eqref{eq:langevin} by itself is ill-defined in the presence of multiplicative noise---one needs to additionally prescribe a suitable integration convention~\cite{gardiner2009stochastic,lau2007state}. It is possible to map Langevin equations from one prescription to another by modifying the deterministic term $f_i$, giving statistically equivalent descriptions of the system (see Appendix A). We can therefore impose the It\^{o} convention \cite{risken1989fpe, gardiner2009stochastic,lau2007state} on all Langevin equations appearing henceforth without loss of generality. Doing so leads to a particularly simple theoretical treatment. 

It should be underlined that we do not require that each and every component of $\mathbf{q}$ follows Langevin dynamics. The upcoming considerations hold for each $q_i$ with $i\leq N$ individually, irrespective of the remaining dynamical details defining the system.
Furthermore, the deterministic terms $f_i$ and noise amplitudes $g_{ij}$ may depend on time $t$ explicitly. In general, Langevin systems may contain multiple Gaussian white noise terms $\eta_j$ that are possibly shared between multiple equations. Then $g_{ij}$ is non-zero for multiple $i$ given any particular $j$. For ease of notation and clarity we focus on cases with single, independent white noises first, rendering the matrix of noise amplitudes $g_{ij}$ diagonal. We examine the influence of multiple noise sources in Sec.~\ref{sec:multinoise}. Finally, our upcoming discussions presume free boundary conditions w.r.t.\ coordinate $q_i$, i.e.\ $q_i \in \mathbb{R}$. Otherwise it is necessary to restrain our analysis to points in phase space with a finite distance to the boundary. We will further elaborate on the importance of this premise later.

Let us consider a temporal discretization up to first order of Eq.~\eqref{eq:langevin} for $k=1$ and rearrange the terms. In the It\^{o} convention, functions $f_i$ and $g_i$ are evaluated at the beginning of the time step and one immediately finds 
\begin{equation}
	\frac{\Delta q_i - f_i(\mathbf{q},t)\Delta t}{g_i(\mathbf{q},t)} =  \Delta W_i(t).
	\label{displacement}
\end{equation}
We assume that $g_i(\mathbf{q},t)\neq 0$ holds and the time step $\Delta t$ is small. The symbol $\Delta q_i$ is a shorthand for the displacement $q_i(t+\Delta t)-q_i(t)$. On the right hand side, $\Delta W_i(t)$ denotes the increments of a Wiener process, distributed according to a Gaussian 
with variance $\Delta t$ and zero mean. Explicitly, its probability density $P(\Delta W_i)$ reads
\begin{equation}
	P(\Delta W_i) = \frac{1}{\sqrt{2\pi \Delta t}}\exp\left( -\frac{\Delta W_{i}^2}{2\Delta t}\right).
	\label{wistribution}
\end{equation}
The displacement $\Delta q_i$ conditioned to an initial state $(\mathbf{q},t)$ similarly follows a Gaussian distribution. This property is exclusive to the It\^{o} convention, which induces a linear relationship between $\Delta q_i$ and $\Delta W_i(t)$ as seen in Eq.~\eqref{displacement}. 
One can now relate the conditional probabilities of two complementary displacements in the coordinate $q_i$ given the same initial phase space position $\mathbf{q}$ at time $t$. Selecting sufficiently small time steps also suppresses the influence of any boundaries that may constrain the possible values of $\Delta q$, returning the displacement probabilities essentially to a Gaussian form. Applying Eqs.~\eqref{displacement} and \eqref{wistribution} yields 
\begin{eqnarray}
	\frac{P(+\Delta q_{i}|\mathbf{q},t)}{P(-\Delta q_{i}|\mathbf{q},t)} 
	= \exp\left[ \frac{2f_i(\mathbf{q},t)}{g_i^2(\mathbf{q},t)} \Delta q_i\right] .
\end{eqnarray}
Remarkably, this expression does no longer explicitly feature the time step $\Delta t$.
To simplify our further discussion we introduce the weighted displacement $\Delta Z_{i}$ obtained from $\Delta q_i$ via a local rescaling:
\begin{equation}
\label{eq: Z_i definition}
	\Delta Z_{i} := \frac{2f_i(\mathbf{q},t)}{g_i^2(\mathbf{q},t)}\Delta q_{i}.
\end{equation}
One should note that the mapping between $\Delta q_i$ and $\Delta Z_i$ is bijective if $\mathbf{q}$ and $t$ are fixed if $f_i(\mathbf{q},t)$ does not vanish.
Since the weighted displacement $\Delta Z_i$ depends linearly on $\Delta q_i$, it is also normally distributed. The distribution's mean $\mu$ is closely connected to its variance $\sigma^2$ via $\sigma^2=2\mu= 4f_i(\mathbf{q},t)/g_i^2(\mathbf{q},t)$.
This condition is necessary and sufficient for a Gaussian to satisfy the symmetry relation
\begin{equation}
	\frac{P(+\Delta Z_{i}|\mathbf{q},t)}{P(-\Delta Z_{i}|\mathbf{q},t)} =
	\exp\left[ \Delta Z_{i} \right].
	\label{eq:flucZ}
\end{equation}
Equation \eqref{eq:flucZ} exhibits the typical form of detailed fluctuation theorems (DFTs), constraining the related probability density by an exponential symmetry and holds to lowest order in $\Delta t$. We can interpret this result as a set of point-wise DFTs, one relation for each of the $N$ Langevin equations.
Because this equality is satisfied for every point in phase space where $g_i$ does not vanish, it seems intuitive that one would retain this symmetry for distributions of the sum of such values as well. This allows us to extend the applicability of Eq.~\eqref{eq:flucZ} to arbitrary sets of initial points and times $Q = \{(\mathbf{q}_j,t_j)\}$, selected independently from the observed displacements $\Delta Z_{i,j}= \Delta Z_i(\mathbf{q}_j,t_j)$. We then define the cumulative weighted displacement $Z_i$ associated with with set $Q$ as
\begin{equation}
\label{eq: Z quantity}
    Z_i := \sum_j \Delta Z_{i}(\mathbf{q}_j,t_j).
\end{equation}
In contrast to most fluctuation theorems known from stochastic thermodynamics, our quantity of interest $Z_i$ is not inherently defined as a functional over continuous trajectories. Rather, it represents a random variable conditioned to a discrete set of initial points $Q$. One can easily derive an extension of Eq.~\eqref{eq: Z_i definition} via convolution of the local probability densities, yielding 
\begin{eqnarray}
	\frac{P(+Z_i|Q)}{P(-Z_i|Q)} = \exp\left[ Z_i \right].
	\label{fluctheo}
\end{eqnarray}
Inductive reasoning allows to prove this conjecture for arbitrary $P(\Delta Z_i|\mathbf{q})$ if they share a common exponential symmetry. Yet this particular problem can be simplified even further by noting that the involved transition probabilities are all Gaussian. Based on the well-known convolution behaviour of normal distributions, we immediately find that the relation connecting mean and variance still holds
\begin{equation}
	\sigma^2(Q) = \sum_j \sigma^2(\mathbf{q}_j,t_j) = \sum_j 2\mu(\mathbf{q}_j,t_j) = 2\mu(Q).
\end{equation}
Equipped with the retained Gaussianity of $P(Z_i|Q)$, this directly implies Eq.~\eqref{fluctheo} for any fixed set of initial points $Q$ as desired. However, the validity of our reasoning hinges on the statistical independence of the initial-state--displacement pairs entering $Z_i$---a premise naturally violated by successive, time-correlated $\Delta Z_i$ taken from a trajectory. 
Fortunately, this apparent limitation in the potential selection or generation methods for $Q$ is easily remedied by drawing random displacements from a long trajectory or large ensemble of trajectory segments, rather than correlated ones. Such an approach fetches initial-state--displacement pairs independently from the marginal probability density $P(\Delta Z_i,\mathbf{q}_j,t_j)$ of the full trajectory distribution, without reference to any $\Delta Z_i(\mathbf{q}_k, t_k)$ with $k\neq j$. The transition from the fully-detailed trajectory picture to adequate marginal distributions is discussed more thoroughly in Sec.~\ref{sec:trajCorrelations}. 
For now, we proceed to further generalize the DFT to cases where the set of initial points $Q$ itself is generated as an independent random variable associated with an arbitrary distribution $P(Q)$. Rather than a convolution we have to deal with a mixture of Gaussians, satisfying
\begin{eqnarray}
\label{eq: general WNDFT}
	P(Z_i) &=& \int dQ\;  P(Z_i|Q)P(Q) = \nonumber \\
	&=& \int dQ\; e^{Z_i} P(-Z_i|Q)P(Q) = \nonumber \\
	& =& e^{Z_i} P(-Z_i).
\end{eqnarray}
Here, the symbol $dQ$ is a shorthand for $\prod_j d\mathbf{q}_j\, dt_j$. Equation~\eqref{eq: general WNDFT} represents the main result of this work. It describes a detailed fluctuation theorem-like relation about the distribution of the quantity $Z_i$ for independently drawn or fixed sets of initial points $Q$.
From Eq.~\eqref{eq: general WNDFT}, we learn that no matter how the (set of) initial conditions $Q$ is distributed, and irrespective whether subsequent evaluations of $Z_i$ even use the same set of starting points, the DFT is satisfied. In practice, this liberates us from the need of sorting displacements according to $\mathbf{q}_j$, for instance obtained from a long experimental trajectory, and gathering a lot of statistics for each initial condition under investigation. It also offers the possibility of arbitrarily prescribing regions of initial points $\mathbf{q}$ to highlight phase space portions of particular interest. In fact, $P(Q)$ is completely arbitrary as long as an appropriate independence condition $P(\Delta Z_{i,j}|Q,\{\Delta Z_{i,k\neq j}\})= P(\Delta Z_{i,j}|q_j,t_j)$ holds. Furthermore, there are no restrictions on the set size of $Q$, causal continuity, existence of stationary states, or time step (if sufficiently small), providing great freedom in how experimental or computational data can be summarized or prepared.

Finally, and in conclusion of our main points, let us see how the DFT directly leads to an integral fluctuation theorem (IFT):
\begin{equation}
	\left\langle e^{-Z_i}\right\rangle = \int_{\mathbb{R}} e^{-Z_i} P(Z_i)dZ_i =\int_{\mathbb{R}} P(-Z_i) dZ_i = 1.
\end{equation}
Here, we have invoked the DFT and used the density's proper normalization. Typically a very large amount of samples is needed to properly evaluate the exponential average if the expectation value of $Z_i$ itself is large. Reducing the time step or the number of addends entering $Z_i$ can help in examining this property but one should avoid $\left\langle Z_i\right\rangle$ becoming so small that the IFT is fulfilled trivially. It is therefore recommended to have a look the distribution of $Z_i$ before assessing statistical significance. This problem is well-known for IFTs in general \cite{jarzynski2006rare}.

\subsection{Multiple Noise Terms}
\label{sec:multinoise}
Let us now consider a discretized Langevin equation in the It\^{o} convention as usual with multiple white noise terms. Following the form of Eq.~\eqref{eq:langevin} we write
\begin{equation}
	\Delta q_i = f_i(\mathbf{q},t)\Delta t + \sum_ j g_{ij}(\mathbf{q},t) \Delta W_{j}.
\end{equation}
The total stochastic displacement results from a sum of rescaled, normally distributed random numbers $g_{ij}\Delta W_j$. The prefactors $g_{ij}(\mathbf{q},t)$ set the variances of individual addends $g_{ij}\Delta W_{ij}$ to $g_{ij}^2\Delta t$.
Their sum is distributed according to the convolution of the individual probability densities---all Gaussians with zero mean. The variance of this new random variable equals the sum of its constituents and one finds the statistically equivalent description
\begin{eqnarray}
	\Delta q_i &=& f_i(\mathbf{q},t)\Delta t + \sqrt{\sum_j g^2_{ij}(\mathbf{q},t) } \Delta W_{qi}, \\
	\dot{q}_i &=& f_i(\mathbf{q},t) + \sqrt{\sum_j g^2_{ij}(\mathbf{q},t) }\eta_{qi}, 
	\label{eq:multiNoise}
\end{eqnarray}
reducing the problem to a single noise term. 

For completeness, one also ought to ask how the fluctuation theorem would be affected if original noise histories were shared between multiple Langevin equations, a case which is frequently discussed in the literature \cite{risken1989fpe}. Fortunately, the answer is quite straightforward: since the DFT emerges for each Langevin equation individually on basis of the normally distributed Wiener process, correlations between equations are not relevant to begin with. Even if we did not know how any of the other coordinates $q_j\neq q_i$ describing the system evolved, as long as the coordinate in question $q_i$ follows a Langevin equation in It\^{o} form, it needs to obey the DFT as formulated. Conflating shared noise histories like in the above expression has no practical impact.

\subsection{Data Processing}
\label{sec:trajCorrelations}

In this section we discuss the preparation and treatment of (discrete) trajectory data in the context of the white-noise DFT. Let us define a discrete trajectory $A$ as a set of states $\mathbf{q}_j=\mathbf{q}(t_j)$ that describe the system at times $t_j$. After generating an initial state according to an arbitrary distribution $P(\mathbf{q}_0)$, the trajectory evolves on basis of the full---and at least partially stochastic---system dynamics. If no stationary state exists it is useful to generate many realizations of this stochastic process, subsequently dubbed trajectory segments. Each segment $A$ is independently initiated via $P(\mathbf{q}_0)$ and then propagated until a selected termination criterion is satisfied---for instance until a preset integration time has elapsed or $\mathbf{q}$ leaves the desired sampling region. This procedure randomly draws segments $A$ from an unspecified trajectory distribution $P(A=\{\mathbf{q}_j\})$. In principle, $P(A)$ is fully determined by $P(\mathbf{q}_0)$, the underlying dynamics, and the chosen termination criteria.

Given a sufficiently large trajectory sample, drawing random displacements of a coordinate $q_i$ from the realizations of $P(A)$ is indistinguishable from sampling the corresponding marginal distribution directly
\begin{equation}
    P(\Delta q_{i,j}, \mathbf{q}_j, t_j) = \left\langle \delta(\Delta q_i(t_j) - \Delta q_i )\delta(\mathbf{q}_j-\mathbf{q}) \right\rangle_A,
\end{equation}
where $\left\langle . \right\rangle_A$ indicates the average over $P(A)$. Repeating this procedure generates arbitrarily large sets of single coordinate displacement $\Delta q_{i,j}:=q_i(t_{j+1})-q_i(t_{j})$ and initial state $\mathbf{q}_j$ pairs without reference to any correlations potentially present at the trajectory level, inducing
\begin{equation}
    P(\{(\Delta q_{i,j}, \mathbf{q}_j, t_j)\}) = \prod_j P(\Delta q_{i,j}, \mathbf{q}_j, t_j).
    \label{eq:marginaldQ}
\end{equation}
This probability density instantly yields $P(\{\Delta Z_j\},Q)$ via a simple coordinate transformation if $q_i$ is governed by a Langevin equation in its It\^{o} representation as usual. The associated conditional probability evidently retains the product form of Eq.~\eqref{eq:marginaldQ} and thus implies 
\begin{equation}
    P(\Delta Z_{i,j}|Q,\{\Delta Z_{i,k\neq j}\})= P(\Delta Z_{i,j}|q_j,t_j).
\end{equation}
This condition is sufficient to prove the DFT's validity for $Z_j = \sum_j \Delta Z_{i,j}$ via convolution as mentioned in the context of Eq.~\eqref{eq: general WNDFT}. After ensuring the independence of all $\Delta Z_{i,j}$, the application of the DFT requires no further knowledge about any of the distributions discussed here. 

While the delineated line of reasoning adequately demonstrates the theoretical usability of trajectory data to generate independent displacement sets, it also inherently assumes the availability of enough data to emulate the marginal distribution of Eq.~\eqref{eq:marginaldQ}. Yet, whenever a system consists of many degrees of freedom $\mathbf{q}$, acquisition of the necessary amount of statistics to this end can become a daunting task. Fortunately, generating a representative sample of the marginal distribution is a sufficient condition, but not a necessary one: most degrees of freedom enter the white-noise DFT only through the local weight $w_i(\mathbf{q},t):=2f_i(\mathbf{q},t)/g_i^2(\mathbf{q},t)$ as associated with coordinate $q_i$. Therefore, obtaining a representative sample for the marginal distribution of the weight along with the displacement is sufficient, i.e.\ $P(\Delta q_i, w_i)$ or $P(\Delta Z_i, w_i)$. In other words, the DFT for any $q_i$ is only concerned with the distribution of two scalars irrespective of the system dimensions, leading to much less stringent data requirements in practice. Imposing stricter termination criteria on the trajectory segments may facilitate data collection even further if desired. Finally, we wish to point out that the idea presented in this section easily generalizes to samples obtained from single long trajectories---for instance by splitting the data into segments with a given duration. The distribution of initial states $P(\mathbf{q}_0)$ for the virtual trajectory segments then approximately corresponds to the stationary phase space density derived from an infinitely long trajectory. This holds true even for non-ergodic systems. Alternatively, one could formulate effectively the same argument without reference to the trajectory ensemble if a stationary state is available to give a marginal distribution $P(\Delta Z_i, w_i)$. The numerical examples covered in the upcoming sections showcase both single- and multiple-trajectory sampling.

\section{Overdamped Systems}
\label{Sec:Overdamped_Systems}

Let us now move on to some practical examples, starting with a setup frequently investigated in the context of stochastic thermodynamics: a Brownian particle in a potential $U(x,\lambda)$ with an external time-dependent protocol $\lambda(t)$. Applied throughout the interval $t\in [t_1,t_2]$, the protocol begins with a system in equilibrium given $\lambda_1$ and ends at $\lambda_2$. 
We will use this example to illustrate some prominent differences to FTs based on time-reversal, in particular the Crooks theorem \cite{crooks1999entropy}. For simplicity we first examine a one-dimensional system governed by the position's equation of motion
\begin{equation}
	\dot{x} = -\frac{1}{m\gamma} \frac{\partial U(x,\lambda)}{\partial x} + \sqrt{\frac{2k_{\text{B}}T}{m\gamma}} \eta(t).
\end{equation}
The gradient of the potential $U(x, \lambda)$ provides the deterministic force, whereas the diffusivity results from the environmental temperature $T$, the particle mass $m$, and the damping coefficient $\gamma$. 
The weighted displacements in this setup assume the form
\begin{equation}
\label{eq: pseudo potential energy difference}
	\Delta Z_x =
	 -\frac{1}{k_{\text{B}}T}  \frac{\partial U}{\partial x}  \Delta x.
\end{equation}
According to conventional definitions in this field \cite{jarzynski1997nonequilibrium, seifert2012stochastic, sekimoto_stochastic2010}, variations in the energy due to a change in the internal degree(s) of freedom are associated with the heat $Q$. Energy changes caused by the protocol are instead summarized into the work $W$. Crooks has shown that the ratio between the probabilities of performing an amount of work $W$ in the forward trajectory and an amount $-W$ in the backward trajectory is given by $\exp[(W-\Delta F)/k_{\text{B}}T]$ if the dynamics respect detailed balance \cite{crooks1999entropy}. This law holds for forward and backward processes starting in equilibrium states whose free energies are $F(\lambda_1)$ and $F(\lambda_2)$ respectively, with $\Delta F = F(\lambda_2) - F(\lambda_1)$. 

However, in the context of the white-noise DFT of Eq.~\eqref{eq: general WNDFT} the stochastic increments must be interpreted in an It\^{o} sense. Equation~\eqref{eq: pseudo potential energy difference}, in general, deviates from the potential energy difference between starting and final position. Therefore $\Delta Z_x$ corresponds to neither heat nor work, but contains additional terms not covered by conventionally relevant thermodynamic measures. That being said, the conditions for the noise-based DFT of Eq.~\eqref{eq: general WNDFT} are more easily fulfilled in practice.
Besides individually holding for each and every point in phase space with finite diffusivity, the noise-symmetry DFT provides a separate relation for each spatial coordinate in higher-dimensional systems. Furthermore, Langevin dynamics abide by this symmetry even if no stationary state exists, directly rendering detailed balance irrelevant as well. 
The DFT can also deal with systems that feature coordinates evolving in a non-Markovian fashion. The removal of these typical requirements appearing in the derivation of time-reversal FTs equips our theorem with a wide range of applicability. For instance, non-conservative forces $\mathcal{F}$ in arbitrary dimensions modify our overdamped equation of motion in the following way
\begin{equation}
	\dot{x}_i = -\frac{1}{m\gamma} \frac{\partial U(\mathbf{x},t)}{\partial x_i}  + \frac{\mathcal{F}_i(\mathbf{x},t)}{m\gamma} +\sqrt{\frac{2k_{\text{B}}T}{m\gamma}} \eta_i(t).
	\label{eq:ncWork}
\end{equation}
The associated quantity 
\begin{equation}
    \Delta Z_i = \frac{1}{k_{\text{B}}T}\left(  \mathcal{F}_i-\frac{\partial U}{\partial x_i}  \right) \Delta x
    \label{eq:nconsZ}
\end{equation}
then includes an additional term that resembles the infinitesimal work performed by the non-conservative force, but again cannot be interpreted as such due to the It\^{o} convention. However, the theorem can be applied without any special treatment or additional considerations involved.

\subsection{Linear Non-Conservative Forces}

\begin{figure}[t!]		
	\centering
	\includegraphics[width=.85\linewidth]{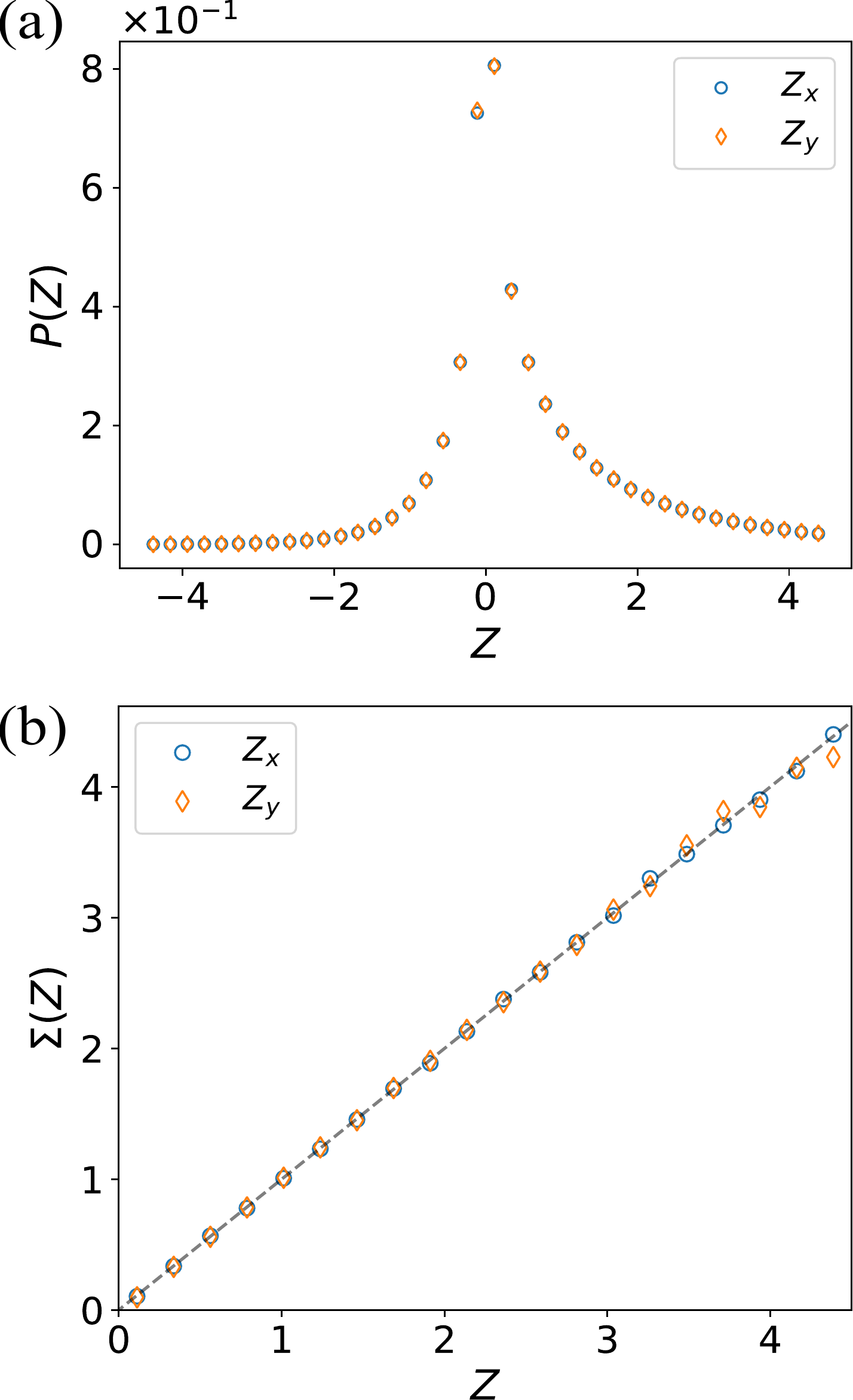}
	\caption{\textbf{Non-conservative system.} A non-conservative force acts on an overdamped particle in a two-dimensional space. The non-conservative driving becomes stronger as the particle moves outward. The particle's indefinite escape is avoided by resetting the trajectory to the origin once the distance to the origin is greater than $s_{m} = 10\sqrt{k_\text{B} T/a} $.  
	Quantifying time in terms of a characteristic physical timescale $t_0 = \sqrt{m/a}$, the damping is given by $\gamma = 100t_0^{-1}$. 
	(\textbf{a}) Histograms of the cumulative weighted position increments $Z$. Each value of $Z$ is accumulated over $50$ uncorrelated displacements with time step $\Delta t = 0.1t_0$.
    The histograms associated with the position-displacements in the $x$- and $y$-direction coincide due to the system's symmetry. (\textbf{b}) Logarithmic probability ratio $\Sigma(Z)$. The results are in excellent agreement with the theoretical reference curve (dashed line).}
	\label{fig:centrifuge}
\end{figure}

As a practical example for a system to our knowledge not already covered by other DFTs, let us direct our attention to a particular two-dimensional system. Specifically, we discuss an overdamped particle subjected to a linear non-conservative force. The field lines of the non-conservative force are circular, driving the particle to revolve around the centre. Due to this external force, energy is continuously fed into the system and subsequently dissipated. The particle's dynamics is governed by the equations
\begin{subequations}
\begin{eqnarray}
    \dot{x} &=&  - \frac{ay}{m\gamma} +\sqrt{\frac{2k_\text{B} T}{m\gamma}}\eta_x, \\
	\dot{y} &=&  + \frac{ax}{m\gamma} +\sqrt{\frac{2k_\text{B} T}{m\gamma}}\eta_y,
\end{eqnarray}
\end{subequations}
where $x$ and $y$ constitute the particle coordinates and $a$ sets the external force's strength. 
The deterministic forces increase in magnitude as the particle moves away from the origin.
Without the presence of a confining potential the particle can drift outward indefinitely. In other words the system has no stationary state. Our example system falls into the class of systems covered by Eq.~\eqref{eq:ncWork}. Here, we have simulated an ensemble of trajectory segments that are terminated if the particle leaves a selected perimeter around the origin. Trajectories begin at the origin with a velocity drawn from the Maxwell-Boltzmann distribution. Irrespective of this initial condition, the termination criterion, and the non-stationarity, the noise-symmetry DFT must hold. 

Each Langevin equation implies a symmetry relation for the respective weighted position displacements $\Delta Z_{x}$ or $\Delta Z_{y}$ analogous to Eq.~\eqref{eq:nconsZ}, which the computed histograms shown in Fig.~\ref{fig:centrifuge} indeed fulfill. Due to the problem's symmetry, $Z_x$ and $Z_y$ are identically distributed. We quantify the histogram symmetry by looking at the logarithmic probability ratio, henceforth dubbed the symmetry function 
\begin{equation}
\label{eq: sigma}
	\Sigma(Z) := \ln \frac{P(+Z)}{P(-Z)}.
\end{equation}
This measure is identical to $Z$ itself if the noise-symmetry DFT is satisfied, becoming a linear function through the origin with unity slope. The lower panel of Fig.~\ref{fig:centrifuge} compares the computed $\Sigma(Z)$-curve against this theoretical prediction, demonstrating excellent consistency.

\section{Underdamped Systems}
\label{Sec:Underdamped_Systems}

\begin{figure}[t!]	
	\centering
	\includegraphics[width=.85\linewidth]{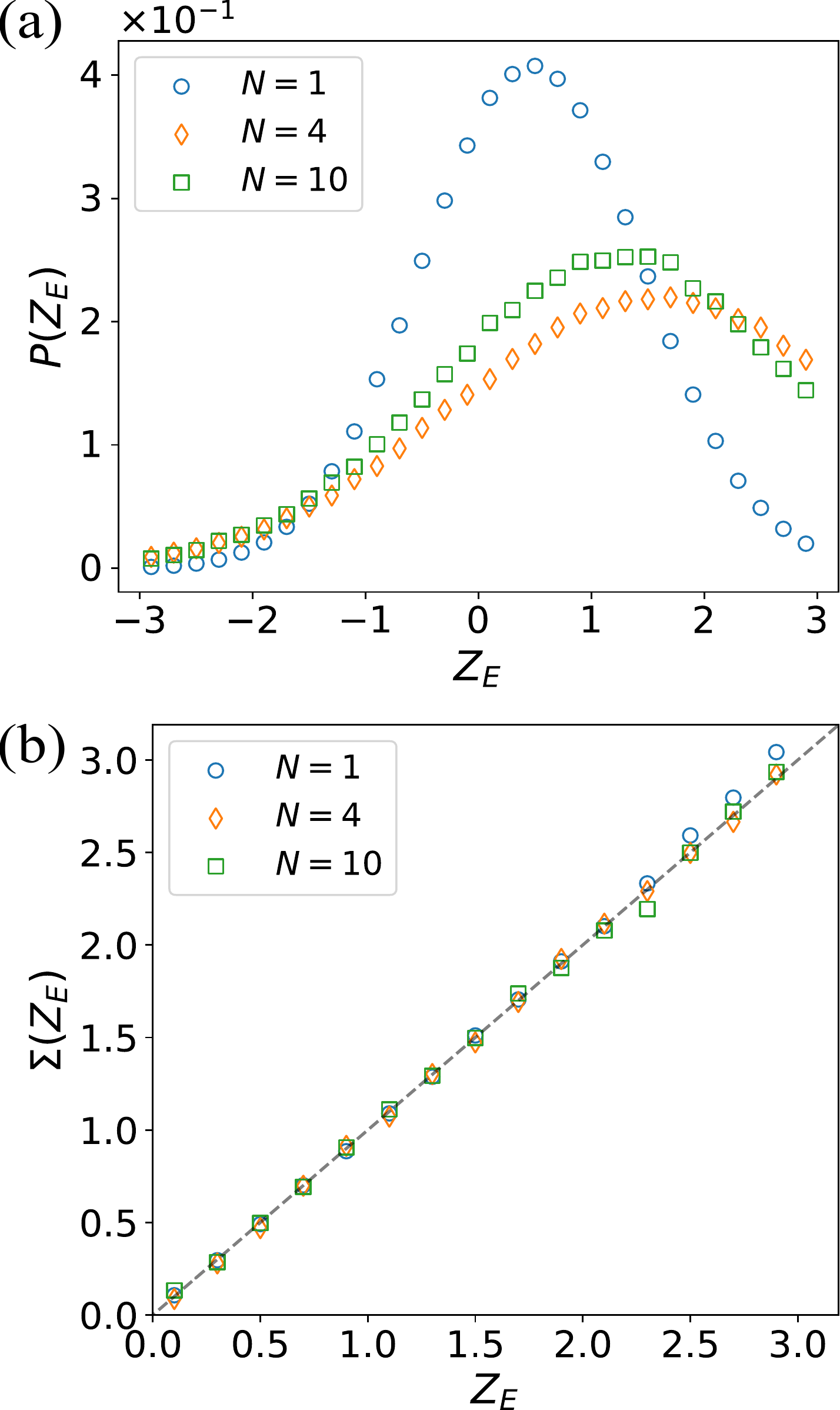}
	\caption{\textbf{White-noise DFT of the total energy in a conservative system.} A set of non-interacting particles is trapped in a harmonic potential. (\textbf{a}) Histograms for three different particle numbers $N=1,4,10$ computed by tracing the total kinetic and potential energy. Each realization of $Z_E$ consists of 250 increments $\Delta Z_E$ drawn from a long trajectory. The parameters of the simulation read $\Delta t = \num{2e-3}t_0$ and $\gamma = t_0^{-1}$, given in units of $t_0=\sqrt{m/k}$. (\textbf{b}) Logarithmic probability ratio $\Sigma(Z_E)$ as given by Eq.~\eqref{eq: sigma}. The histograms respect the DFT for all particle numbers $N$ considered.}
	\label{fig:energy}
\end{figure}

In this section, we will apply the white-noise DFT to simulated trajectories of underdamped systems. We use our results to explore the versatility of the white-noise DFT for systems that feature multiplicative noise histories, non-Markovian dynamics, and variables not invariant under time-reversal like the velocity.

\subsection{Energy Diffusion}

Particles subjected to conservative forces in a thermal bath constitute a particularly important and ubiquitous type of Langevin dynamics in physics. Subsequently, the index $i\in \{1,\dots,N\}$ numbers the coordinates $x_i$ and velocities $v_i$ of individual particles in any particular spatial dimension. The evolution of the degrees of freedom is described by
\begin{eqnarray}
    \dot{x}_i &=& v_i, \\
	m\dot{v}_i &=& -\frac{\partial U(\mathbf{x})}{\partial x_i}  - m\gamma v_i +\sqrt{2m\gamma k_\text{B} T}\eta_i.
\end{eqnarray}
The potential $U(\mathbf{x})$ only depends on the particle coordinates $x_i$, but is otherwise arbitrary. Furthermore, energy is dissipated to the environment through a linear damping force proportional to the particle velocity $v_i$ in each direction. Each of these relations results in a separate DFT for the weighted velocity displacements
\begin{equation}
	\Delta Z_i = -\frac{1}{\gamma k_{\text{B}}T} \left(  \frac{\partial U(\mathbf{x})}{\partial x_i} + m\gamma v_i\right)\Delta v_i .
\end{equation}
However, it is also possible to consider the symmetries of composite measures, i.e.\ functions of the particle positions and velocities. The total energy of the system may represent the most prominent example in this context. It can be written as the sum of the potential and the total kinetic energy $K$,
\begin{equation}
	E = U(\mathbf{x})+ \sum_{i} \frac{mv^2_i}{2}.
\end{equation}
A Langevin equation for the total energy $E$ with the appropriate integration convention is easily derived by invoking It\^{o}'s lemma \cite{gardiner2009stochastic}:
\begin{eqnarray}
\label{eq: energy evolution}
	\frac{dE}{dt} &=& \sum_{i}\left( \gamma k_\text{B} T -m\gamma v^2_i + v_i \sqrt{2m\gamma k_\text{B} T}\eta_i  \right) \nonumber\\
	&=& N\gamma  k_\text{B} T-2\gamma K + \sqrt{4\gamma Kk_\text{B} T }\eta_E.
\end{eqnarray}
Here, we have recycled the reasoning used in Eq.~\eqref{eq:multiNoise} to summarize multiple Gaussian white noises into a single one. As we can see from Eq.~\eqref{eq: energy evolution}, the changes of the total energy only depend on the dissipation and the fluctuations due to the environment. In the absence of environmental forces, the potential energy $U(\mathbf{x})$ could only be converted into kinetic energy and vice versa, leaving the total energy constant.
One immediately obtains the weighted energy displacements
\begin{equation}
    \Delta Z_E = \frac{Nk_\text{B} T-2K}{2k_\text{B} T K}\Delta E.
\end{equation}
Irrespective of any details concerning the potentially very complex interactions in the system, energy displacements have to obey the simple law
\begin{equation}
	\frac{P(+\Delta Z_E)}{P(-\Delta Z_E)} = \exp \left( {\frac{Nk_\text{B} T-2K}{2k_\text{B} T K} \Delta E} \right) .
	\label{eq:energy}
\end{equation}
There are several points of interest here: contrary to our previous examples that tacitly investigated coordinates with free boundary conditions, the system energy often exhibits a lower bound. The presence of non-trivial boundary conditions influences the displacement probabilities in the immediate vicinity of boundaries, distorting their Gaussian statistics. Despite inducing discrepancies in the DFT, one can always find a sufficiently small time step to recover its validity if the distance to the boundary is finite. Additionally, since the DFT symmetry applies locally, one can arbitrarily prescribe the regions in parameter space from which initial points are drawn, e.g.\ the exclusive sampling of displacements with starting energies greater than any desired threshold. In practice, the influence of boundaries on the theorem can frequently be neglected or eliminated by choosing an adequate time discretization and sampling region. Furthermore, it is noteworthy that the energy's evolution features multiplicative noise, i.e.\ a diffusivity that depends on $E$ itself via $K$. 

To support the validity of the white-noise DFT for this multiplicative noise case, we consider a system of $N$ non-interacting particles in one dimension confined by a harmonic potential $U(x) = kx^2/2$. After generating a long trajectory as test data, we have evaluated the DFT according to Eq.~\eqref{eq:energy}, sampling 250 randomly drawn $\Delta Z_E$ for each realization of $Z_E$. We repeat our analysis for different numbers of particles $N$, achieving the distributions shown in Fig.~\ref{fig:energy}(a). These distributions differ significantly from one another, yet they all fulfill the white-noise DFT, as we can appreciate from Fig.~\ref{fig:energy}(b).

\subsection{Delayed Feedback}

\begin{figure}[t!]	
	\centering
	\includegraphics[width=.85\linewidth]{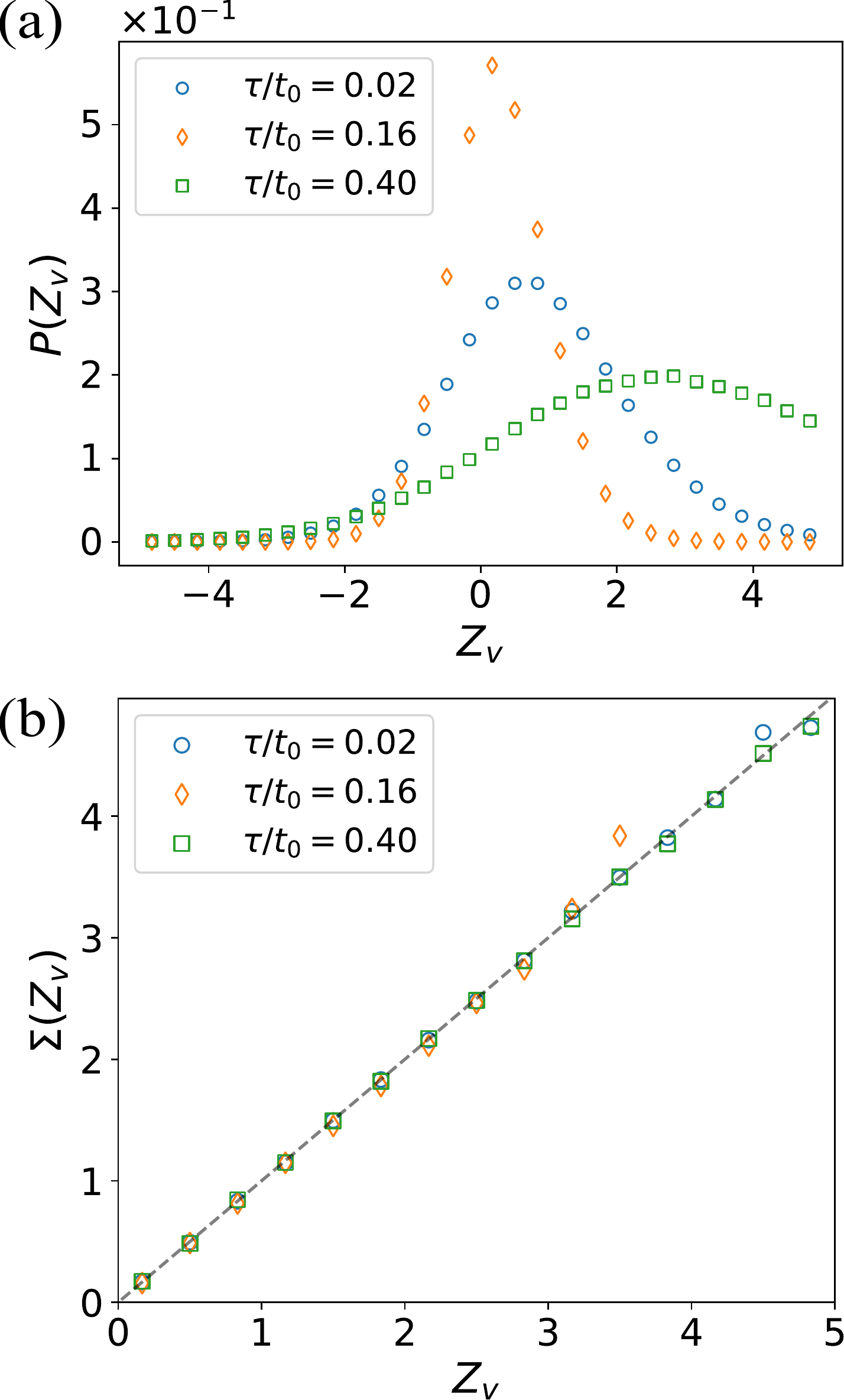}\label{fig:delayB}
	\caption{\textbf{System with feedback delay.} A particle in a harmonic potential is subjected to an additional feedback force that drives it away from the centre in proportion to its position at a previous time $x(t-\tau)$. Despite the dynamics' overall non-Markovian nature, the DFT is still obeyed in the velocity variable. (\textbf{a}) Histograms for three different delay times $\tau/t_0 =0.02,\,0.16,\,0.40$, given in units of the characteristic timescale $t_0=\sqrt{m/k}$. The displacements of quantity $Z_v$ are accumulated over ten time steps of length $\Delta t = \num{e-2} t_0$. Remaining parameters: $\gamma/2\pi =0.02 $, $l=k/2$. (\textbf{b}) Logarithmic probability ratio $\Sigma(Z_v)$ along with the theoretical reference (dashed gray line). All datasets obey the DFT within statistical error.}
	\label{fig:delay}
\end{figure}

In this subsection we investigate an example of non-Markovian dynamics to which the white-noise DFT can be applied.
We consider a one-dimensional harmonic oscillator in a thermal bath that is additionally subjected to an external feedback force $z(t)$. The feedback is proportional to the particle's previous position and applied with a delay of constant size $\tau$, rendering the system non-Markovian as intended. For simplicity, we assume that the feedback force is linear like its conservative counterpart, leading us to the set of equations shown below 
\begin{subequations}
\begin{eqnarray}
	\dot{x} &=& v, \label{eq: position langevin}\\
	z &=& x(t-\tau), \label{eq: feedback langevin} \\
	m\dot{v} &=& -kx + lz +\sqrt{2m\gamma k_\text{B} T}\eta.  \label{eq: velocity Langevin}
\end{eqnarray}
\label{eq:delayLang}
\end{subequations}
The spring constant $k$ is assumed to be positive, whereas the proportionality constant $l$ of the feedback force may be selected arbitrarily. This system constitutes the first example that is not solely represented by Langevin equations. The evolution of the associated density in probability space cannot be described by a single Fokker-Planck equation either. Nonetheless, our symmetry relation remains applicable for the velocity Langevin equation in Eq.~\eqref{eq: velocity Langevin}, as confirmed by the simulation results depicted in Fig.~\ref{fig:delay}. We show histograms for three different delay times $\tau$, ranging from values much shorter than the potential's characteristic oscillation time $2\pi\sqrt{m/k}$ to delays that are in the same order of magnitude. Despite leading to very distinct distributions of the displacement variable $\Delta Z_v$, all of the computed distributions clearly share the predicted symmetry as highlighted in Fig.~\ref{fig:delay}(b). The full evolution of this system depends on a prior point in time and is thus non-Markovian. Furthermore, the delay time may assume arbitrarily large values. For systems such as these no other detailed or even just integral fluctuation theorems exist to our knowledge.

\section{Parameter inference}
\label{Sec:Inference}

Contrary to fluctuation theorems based on time-reversal, the quantities $Z$ appearing in the white-noise DFT are not directly tied to entropy, heat or work. In fact, $Z$ only describes coordinate displacements that are reweighted by the local competition of deterministic and stochastic forces. The weight takes the value $2f_i/g^2_i$ as shown in Eq.~\eqref{eq: Z_i definition}, equal to the ratio of the local deterministic force and diffusivity. Assigning physical meaning to $Z$ beyond this is difficult. 
Yet it is precisely this difference from conventional fluctuation theorems that opens up the large variety of systems it can be applied to. In this section, we will demonstrate how this versatility leads to an efficient and flexible method to infer unknown parameters of a system, and how parameters of an effective/simplified model can be optimized to be as consistent as possible with some reference data.
For notational simplicity we exclusively discuss one-dimensional systems with time-independent forces in the following. Note however that every aspect of our subsequent considerations is easily generalized to time-dependent multi-dimensional systems. 

\subsection{Uniqueness}

Assume we are given a trajectory of a coordinate $q$, which evolves according to a Langevin equation $\dot{q}=f(q)+g(q)\eta$, yet is modelled on the basis of a different one $\dot{q}=f_m(q)+g_m(q)\eta$. Discrepancies of this kind are very common in practice and may stem from experimental uncertainties in one or more parameters or the application of simplified models like the overdamped approximation. The question we want to answer in the following is: how do differences between model and true dynamics quantitatively affect the white-noise DFT? 

Let us first define the model displacement $\Delta Z_m = 2f_m(q)\Delta q/g_m^2(q)$ and deviation factor $r(q)= \Delta Z/\Delta Z_m$. Note that the latter does not depend on $\Delta q$. We restrict our analysis to positions $q$ where neither the model's deterministic term nor the diffusivity vanish, guaranteeing that $r$ exists and is finite. With these definitions and the Gaussianity of $\Delta Z_m$ inherited from $\Delta q$, it is easy to see that the distribution $P(\Delta Z_m)$ satisfies the relation
\begin{equation}
	\frac{P(+\Delta Z_m|q)}{P(-\Delta Z_m|q)} = e^{r(q) \Delta Z_m}.
	\label{eq:modelFluc}
\end{equation}
We can consequently extract the correction factor $r(q)$ from the slope of our local symmetry function $\Sigma(\Delta Z_m)$. Alternatively, since the underlying distribution is still Gaussian one can take advantage of the numerically more stable expression $r(q)=2\mu_m(q)/\sigma_m^2(q)$ instead, where $\mu_m$ and $\sigma_m^2$ designate the mean value and variance of $\Delta Z_m$ respectively. In contrast, the true $\Delta Z$ derived from $f(q)$ and $g(q)$ would fulfill the relation $2\mu(q)/\sigma^2(q)=1$ for all $q$ per construction.  
Any discrepancy between the true dynamics and the model at position $q$ is thus uniquely inferable via $r$ up to a common prefactor in $f(q)$ and $g^2 (q)$. To put it another way: by invoking the equality $r f_m/g^2_m = f/g^2$ we can determine $h(q)f(q)$ and $h(q)g^2(q)$, where $h(q)$ is an arbitrary real number. This multiplicity is a consequence of the DFTs time step independence.
Although the above equation can be employed to extract $f/g^2$ for any initial position, this approach is impractical as one would need to gather sufficient statistics for each $q$. Even if enough data were available, it would be more practical to calculate the average displacement and its variance to obtain $f(q)$ and $g^2(q)$ without the delineated multiplicity. Nonetheless, the DFT can be used in other ways, which we will discuss shortly.
Before doing so, however, it is instructive to investigate the distribution of a sum of model displacements $Z_m=\sum_j \Delta Z_m(q_j)$ taken from different positions in phase space $Q=\{ q_j\}$. As a convolution, the associated distribution is another Gaussian with mean $\mu_m(Q)=\sum_j \mu_m(q_j)$ and variance $\sigma_m^2(Q)= \sum_j \sigma^2_m(q_j)$. Therefore we obtain
\begin{equation}
\label{eq: rm expression}
	r(Q) = \frac{2\mu_m(Q)}{\sigma_m^2(Q)} = 
	\frac{\sum_j r(q_j)\sigma_m^2(q_j)}{\sum_j \sigma_m^2(q_j)}.
\end{equation}
We find that the correction factor of the sum equals an average which is weighted by its constituents' variances. The exponential symmetry is retained even for arbitrary sums of $\Delta Z_m$, provided that one fixes the initial conditions of the displacements. We discuss the implications of dropping this constraint in the next section.

\subsection{Convergence}
\label{sec:convergence}

It is far more convenient to consider the statistics of $\Delta Z_m$ or $Z_m$ without sorting the displacements in a discrete/discretized trajectory w.r.t.\ their initial positions. The corresponding distribution is no longer a Gaussian, but rather a mixture of Gaussians that can be written in the decomposed form
\begin{equation}
	P(\Delta Z_m) = \int dq\;  P(\Delta Z_m|q)P(q),
	\label{eq:mix}
\end{equation}
where $P(q)$ is the probability of observing initial condition $q$ (or a set of initial points) throughout the dataset. As justified in Sec.~\ref{sec:trajCorrelations}, this probability can be defined even if no stationary state exists. The conditional probability $P(\Delta Z|q)$ satisfies Eq.~\eqref{eq:modelFluc}. We proceed to have a look at mixture distribution's symmetry
\begin{equation}
	 \frac{P(+\Delta Z_m)}{P(-\Delta Z_m)}=\frac{\int dq\; e^{r(q)\Delta Z_m }  P(-\Delta Z_m|q) P(q)}{\int dq\; P(-\Delta Z_m|q)P(q)}.
\end{equation}
The above expression takes the form of a conditional expectation value for $e^{r(q)\Delta Z_m }$, constraining the density to $-\Delta Z_m$.
No simple exponential symmetry exists if $r(q)$ is not constant throughout the phase space portion considered. However, it is easy to show that for positive $\Delta Z_m$ the symmetry function $\Sigma(\Delta Z_m)$ is bounded by
\begin{equation}
	\min_q[r(q)] \Delta Z_m \leq \Sigma(\Delta Z_m) \leq \max_q[r(q)] \Delta Z_m.
\end{equation}
For $\Delta Z_m<0$ one obtains a similar relation, but the inequalities are reversed, implying that $\Sigma$ is encased by two linear functions intersecting at the origin. Fortunately, we can drastically improve upon providing simple bounds by considering sums of displacements drawn randomly from the trajectory or dataset. Again, drawing them randomly serves the purpose of removing correlations between subsequent $r(q)$ if applicable, sampling the marginal distribution instead. As the number of terms in the sum is increased, we approach the weighted average of the dataset
\begin{equation}
    R = \dfrac{\int dq\;  r(q) \sigma_m^2(q) P(q) }{\int dq\;  \sigma_m^2(q) P(q) },
\end{equation}
equivalent to the limit of infinite samples in Eq.~\eqref{eq: rm expression}. While the potential minimum and maximum values of $r(Q=\{q_j\})$ coincide with the respective single point values, the distribution becomes sharply peaked around $R$ for a large number of addends. The symmetry function $\Sigma$ then quickly approaches the value $R\Delta Z_m$ for any finite $\Delta Z_m$, converging towards a linear expression again. 
The slope of a linear fit to $\Sigma$ can therefore be seen as an adequate quantifier for the average consistency of a dynamical model with the data. This statement can also be adapted to selected portions of phase space, if desired.

On basis of the delineated properties we propose a simple, yet versatile optimization algorithm consisting of three steps: (1) postulate a model with a set of free parameters $\mathbf{k}=\{ k_i\}$, (2) compute the distribution of randomly added displacements of the trace to be analyzed and extract $\Sigma(Z(\mathbf{k}))$, (3) vary the parameters $k_i$ to minimize the deviation between $\Sigma(Z(\mathbf{k}))$ and the theoretical reference $Z(\mathbf{k})$, repeating step (2) for each test value of $\mathbf{k}$. One can quantify this deviation via the (initial) slope of our symmetry function $R(\mathbf{k})$ or other simple measures.

Our method is closely related to Bayesian parameter inference, as both approaches aim to replicate the expected statistics of Gaussian white noise. Indeed, by extracting the noise history or increments of the Wiener process from the stochastic displacements, one can also optimize the parameters by maximizing the associated likelihood function $L(\mathbf{k})$ \cite{krog2017bayesian}. While the computational effort necessary for both approaches scales linearly with the number of displacements available, they offer different advantages. On a general basis, however, the performances of these two methods are quite similar and they are almost equally versatile.

\subsection{Comparison with Bayesian Inference}
Let us proceed to elaborate upon several important differences between direct Bayesian inference and the DFT-method. The Bayesian inference is able to identify $f$ and $g$ directly as indicated, avoiding the multiplicity of shared prefactors. Yet it does feature some arbitrariness in the form of the selected prior, influencing the final result.
The DFT analysis in turn drops any explicit dependence on the time step, a very sensitive parameter in Bayesian estimates that may itself be subject to uncertainties. Also, whether the incurred multiplicity is actually relevant for the optimization depends on the set of parameters investigated. Only if one of the parameters leads to a modification of $f$ and $g$ that leaves $f/g^2$ invariant is there additional information to be extracted with the Bayesian approach.
The DFT-method additionally offers a direct quantification of how accurate a given set of parameters is via $R$---more so than the value of the likelihood function $L$ in Bayesian estimation. Except for statistical errors, $R$ becomes unity at the position of the optimal parameter set. The procedure can be terminated once the difference to the reference value is smaller than a threshold value $\epsilon$. 

Next, we underline that whenever $f_m$ and $g_m$ are continuous in $k_i$, both $R(\mathbf{k})$ and $L(\mathbf{k})$ depend continuously on all $k_i$ as well, covering many practically relevant cases. In such scenarios, precise knowledge about our target value $R=1$ is valuable in reducing the computational cost expended to reach the optimal parameter set. 
For instance, after finding two parameter sets $\mathbf{k}_1$ and $\mathbf{k}_2$ such that $R(\mathbf{k}_1)<1<R(\mathbf{k}_2)$ holds, one can close in on the optimum with a bisection algorithm. This procedure reaches a minimum far faster than the Markov chain Monte Carlo (MCMC) approaches typically used with Bayesian inference. The possibility of using bisection with similar Bayes methods is not precluded of course, but necessitates the identification of a concave (boundary) region of $L(\mathbf{k})$. Locating such regions is more computationally expensive as it inherently requires a greater set of likelihood evaluations, especially in higher-dimensional parameter spaces.  
Even if bisection was used in the Bayesian inference after this preparation step, knowledge about $R$ can be utilized to accelerate convergence more effectively: the slope provides additional information to gauge the distance to the optimum and potentially lower the number of iterations necessary to reach it. 
Nevertheless, although the total computational cost scales identically, individual iterations are slightly more expensive in the DFT-method due to the extraction of the symmetry function $\Sigma(Z(\mathbf{k}))$ from the histograms and its linear fit. One can circumvent this step only under the condition that the set of initial points $Q$ is fixed, at which point $R$ is efficiently inferred from the usual expression $2\mu(Q)/\sigma^2(Q)$.   

Lastly, the methods significantly differ in their treatment of false positives---that is local maxima of $L(\mathbf{k})$ or $R=1$ parameter sets that are incompatible with the system's true dynamics, even as a simplification.
Broadly speaking, one does not necessarily reach $R=1$ for any set $\mathbf{k}$ in cases of model incompatibility, but can generally still locate a maximum in the likelihood function. A less efficient optimization in the sense that one searches a minimum of $|R-1|$ remains possible, but the DFT-based algorithm communicates such discrepancies more directly. In the presence of multiple Langevin equations the parameter set must even satisfy several independent DFT relations, making it the much more stringent requirement. Moreover, dynamics not governed by Gaussian white noise generally do not lead to linear $\Sigma$-curves, providing an additional tool for error assessment.
Nevertheless, false positives can in principle occur in either algorithm. One strategy for their identification consists in examining one or more phase space subregions of initial states. Checking for continued consistency is once more computationally cheaper for the DFT-method, as one only needs to find the $R$ value for the test parameter set and see whether the selected termination criterion $|R-1|<\epsilon$ is still satisfied. The Bayes approach necessitates a number of evaluations in the immediate vicinity again to confirm that a local minimum remains approximately at the tested position.

\subsection{Example: Delayed System}

\begin{figure}[t!]		
	\centering
	\includegraphics[width=.85\linewidth]{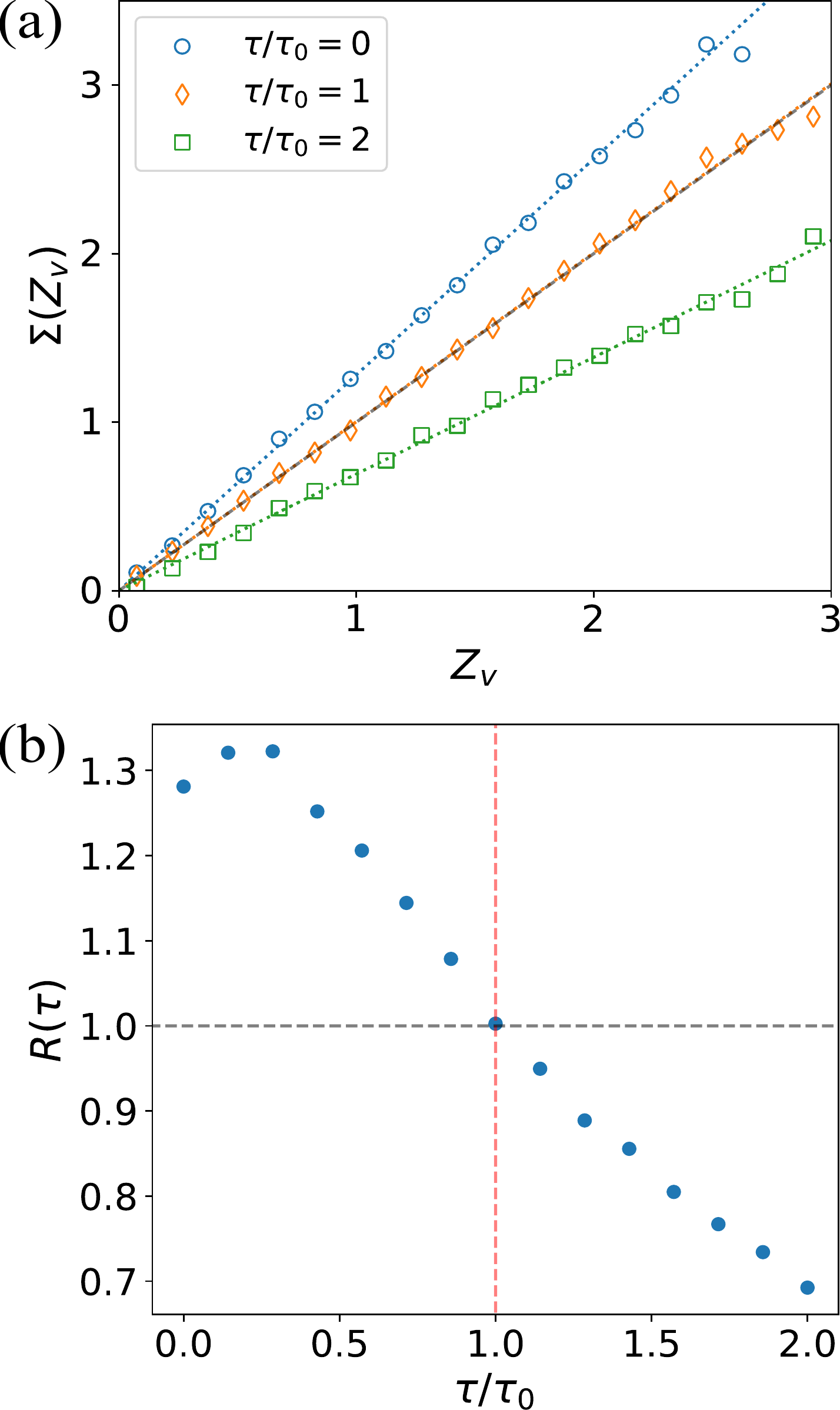}
	\caption{\textbf{Inference of delay time $\tau$.} (\textbf{a}) Logarithmic probability ratio $\Sigma$ of the delayed system described by Eq.~\eqref{eq:delayLang} for three different test delay times, equal to $\tau/\tau_0=0,1,2$. Only the true delay time $\tau=\tau_0$, used for the generation of the trajectory data, results in a graph consistent with the theoretical reference $R(\tau) = 1$ (shown by the black dashed line). Irrespective of the test value $\tau$, the initial behaviour of $\Sigma$ is well-described by a linear approximation. Here, we have analyzed the $\tau_0/t_0=0.16$ data of Fig.~\ref{fig:delay}. (\textbf{b}) Initial slope $R$ of $\Sigma$ as a function of the delay time $\tau$. The theoretically predicted slope of unity is found precisely at the true delay time $\tau_0$.}
	\label{fig:optim}
\end{figure}

Let us revisit one of our previous examples---the harmonic oscillator with delayed feedback. The fluctuation theorem offers an approach to infer most of the system's parameters, for instance $k,l$ and $\tau$ from Eq.~\eqref{eq:delayLang}. Here, we check the accuracy of our optimization algorithm by extracting the delay time $\tau$ that prescribes the degree of non-Markovianity of the system, assuming that the remaining parameters are known beforehand. The displacement weights $f(x,v,z)/D$ explicitly depend on the delay time, implying that remnant discrepancies to the theoretical reference will only disappear at $\tau=\tau_0$ for arbitrary phase space regions and infinitesimal time steps. We generate trajectories with a given delay time $\tau_0$ and subsequently compute the $\tau$-dependent weighted displacement distributions $P(Z,\tau)$. In this artificial example we have computed the displacement statistics of $\tau$-values on a uniformly-spaced grid ranging from zero delay to $2\tau_0$.

Figure~\ref{fig:optim}(a) demonstrates how the symmetry of the extracted distributions can deviate from the theoretical prediction if the test value for $\tau$ strays significantly from $\tau_0$. Nonetheless, 
the logarithmic probability ratio $\Sigma(Z_v)$ remains well-approximated by linear fitting functions: as predicted in Sec.~\ref{sec:convergence}, increasing the number of addends induces convergence to linear $\Sigma(Z_v)$-functions even for deviating models. Here we use the same number of displacements per $Z_v$-value as in Fig.~\ref{fig:delay}. 

Figure~\ref{fig:optim}(b) showcases the extracted initial slopes $R$ as a function of the delay time $\tau$. Corresponding to the average deviation factor $r(x,v,z)$ of the sample, we find that the theoretically predicted slope of $R=1$ is only obtained for the true delay time $\tau_0$. Any remnant discrepancies at this point are negligible and caused primarily by statistical uncertainties.

\section{Signal Contamination}
\label{sec:signalContamination}

\begin{figure}[t!]		
	\centering
	\includegraphics[width=.85\linewidth]{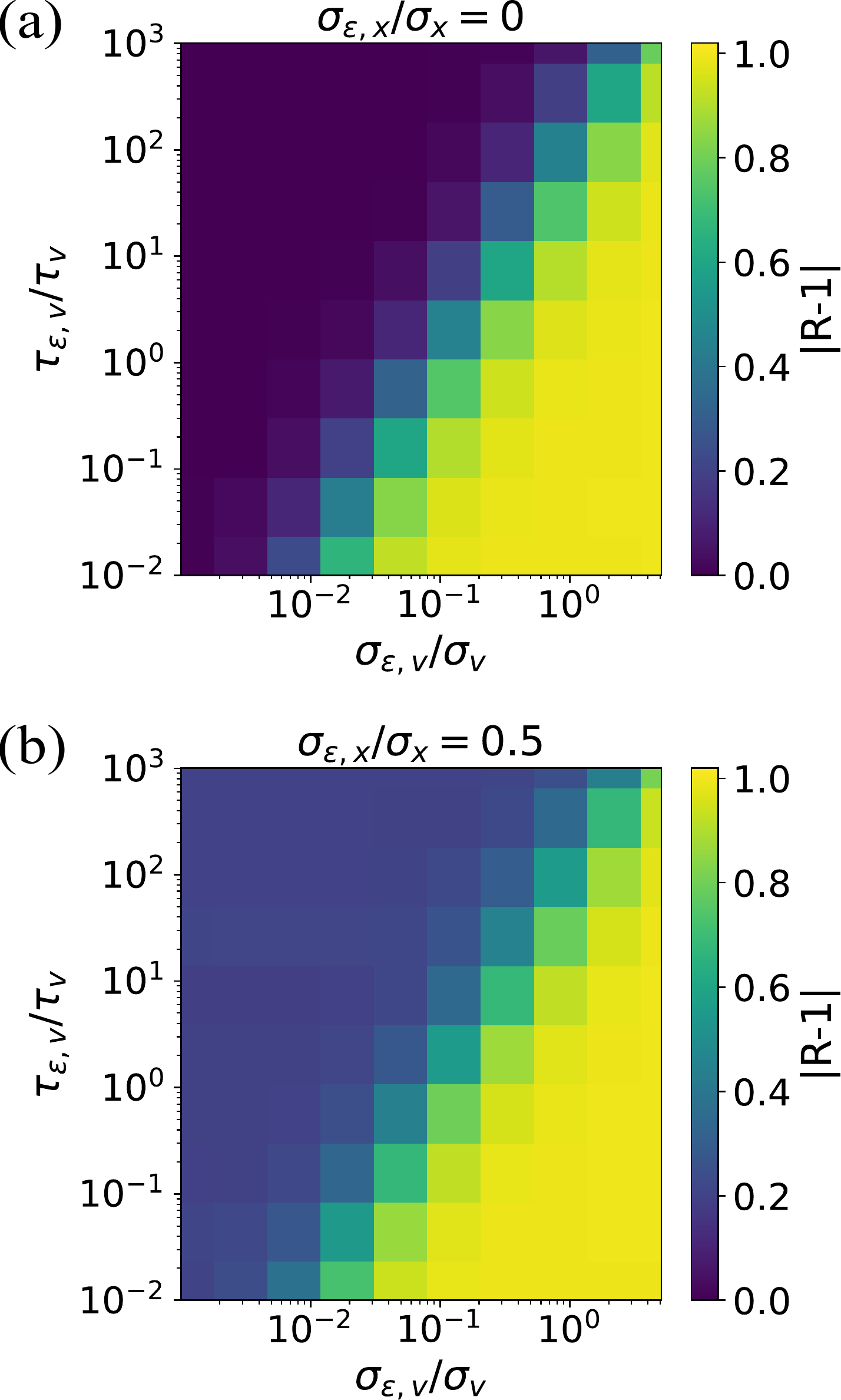}
	\caption{\textbf{Slope of $\Sigma(Z_x)$ as a function of the signal noise parameters.} 
	The heatmaps showcase the deviation w.r.t.\ the expected slope $R=1$ as a function of the noise's magnitude $\sigma_{\epsilon, v}$ and correlation time $\tau_{\epsilon,v}$ as it contributes to the velocity signal. The respective values are given in terms of the unbiased velocity's standard deviation $\sigma_v$ and decorrelation time $\tau_v = 1/\gamma$.
	The two panels only differ in the magnitude of the position error, equalling $\sigma_{\epsilon, x} = 0 $ in (\textbf{a})~and $\sigma_{\epsilon, x} = \sigma_x/2$ in (\textbf{b}).} \label{fig:noiseMesh}
\end{figure}

The practical relevance of the white-noise DFT lies in the ability to infer parameters from simulated and experimental trajectories. Especially in the latter case it may help in extracting otherwise elusive information from the short-time evolution of the system. Yet, this application inherently demands the availability of sufficiently accurate displacement data. Whilst our inference method can aid in identifying accurate dynamical descriptions and extracting adequate parameter sets, measurement noise remains a fundamentally unavoidable source of errors. It therefore merits a more thorough qualitative discussion.

We subsequently investigate the sensitivity of the DFT with respect to noise as a function of its standard deviation $\sigma_\epsilon$ and correlation time $\tau_\epsilon$. 
Specifically, we examine an underdamped harmonic oscillator in a thermal bath described by
\begin{subequations}
	\begin{eqnarray}
		\dot{x} & =& v, \\  
		\dot{v} & =& -\gamma v - \omega_0^2 x +\sqrt{2D}\eta.
	\end{eqnarray}
\end{subequations}  
Its position and velocity are subject to Gaussian measurement uncertainties $\epsilon$, generating signals $x'=x+\epsilon_x$ and $v'=v+\epsilon_v$. 
We base the evolution of the signal errors on Ornstein-Uhlenbeck processes
\begin{subequations}
	\begin{eqnarray}
		\dot{\epsilon}_x &=& - \frac{\epsilon_x}{\tau_{\epsilon, x}}  +\sqrt{\frac{2\sigma_{\epsilon, x}^2}{\tau_{\epsilon, x}}} \eta_{\epsilon, x}, \\
		\dot{\epsilon}_v &=& - \frac{\epsilon_v}{\tau_{\epsilon, v}}  +\sqrt{\frac{2\sigma_{\epsilon, v}^2}{\tau_{\epsilon, v}}}  \eta_{\epsilon, v},
	\end{eqnarray}
\end{subequations}  
allowing us to freely prescribe the noise magnitudes $\sigma_{\epsilon, x}$, $\sigma_{\epsilon, v}$ and correlation times $\tau_{\epsilon, x}$, $\tau_{\epsilon, v}$. The Ornstein-Uhlenbeck toy model further offers the unique advantage of ensuring that even the corrupted signal exhibits Gaussian transition probabilities, retaining the linear DFT curves as discussed in the main text.
 
In a real experiment the velocity signal would be typically extracted from the position via finite difference methods or fitting procedures. For greater simplicity and generality we make no particular assumption on the signal retrieval method. Here we consider the errors $\epsilon_x$ and $\epsilon_v$ to originate from uncorrelated red noises. Without any information about the error's value, the signals $x'$ and $v'$ become substitutes for the real position $x$ and velocity $v$ and we evaluate the DFT of the corresponding velocity Langevin equation, i.e.\ for the quantity
\begin{equation}
	\Delta Z_v' := \frac{-\gamma v' -\omega^2 x'}{D} \Delta v' .
\end{equation}  
Figure \ref{fig:noiseMesh} outlines the impact of the signal velocity's decorrelation time and noise magnitude. As expected, the slope of the logarithmic probability ratio curves returns to unity if the signal errors become negligibly small. Deviations appear and become evermore prominent as we increase in noise intensity. However, longer noise correlation results in a slower onset of this effect, partially mitigating the impact of the signal error. This observation is likely owed to the fact that errors in the DFT originate from two sources: the phase space point before and after the displacement. The ensemble of measured initial points remains invariant under changes in $\tau_\epsilon$, however the displacement itself is deeply sensitive w.r.t.\ this parameter. Longer correlation times diminish the signal error's change between time steps and therefore allow for the extraction of a more accurate estimate of the velocity difference---but not of the velocity itself. Although white noise finds more frequent application as an error model in practice, it is precisely this correlation that leads to more faithful $Z_v'$-histograms.  
Nonetheless, as one steps further into the regime of strong corruption, we seemingly encounter a threshold beyond which the DFT breaks down irrespective of $\tau_\epsilon$. This occurs at a signal-to-noise ratio of approximately one, i.e.\ when the signal corruption starts to overshadow the physical movement completely. 

Although qualitatively unchanged, this effect is also strongly sensitive to changes in the time step used throughout the analysis. Small time steps lead to reduced displacement magnitudes as well. Therefore fast fluctuations caused by signal errors may bury the physical information more easily. This circumstance imposes a lower limit on the time step size for non-continuous noise histories. Naturally, a more general limitation restricts the use of larger $\Delta t$: the validity of the DFT is inherently tied to short timescales at which the average drift and diffusivity reliably follow from the linear and quadratic displacement. This newly added complication of a lower bound due to measurement noise can severely reduce the range of viable $\Delta t$. Using extremely small time steps can only be recommended for continuous noise histories, where most benefits are achieved only after crossing the correlation time threshold. Applications that seek to take advantage of the white-noise DFT need to mitigate this issue more effectively, thus a high signal-to-noise ratio becomes indispensable.

\section{Discussion}
\label{Sec:Conclusions}

In this work we have introduced a class of fluctuation theorem-like symmetries applicable to all kinds of Langevin dynamics. Contrary to established fluctuation theorems central to the field of stochastic thermodynamics, these are not based on any form of time-reversal, but rather on the statistics of the underlying Gaussian white noise. Consequently, the measures appearing in the theorems are not associated with entropy or other typical thermodynamic quantities either. They instead relate to short-time displacements reweighted by the local competition between deterministic and stochastic forces. 
The noise-symmetry DFTs frequently hold even for systems that remain elusive to their time-reversal counterparts: they neither require any knowledge about stationary states nor the existence thereof and even apply to non-Markovian systems that feature long-time memory. The only requirement is that Langevin equations describe the evolution of one or more degrees of freedom. 
This versatility leads to a powerful time-step--independent inference algorithm that may be utilized to extract various parameters from computational and experimental trajectory data. The method is inherently connected to Bayesian inference, but has built-in termination criteria and features information that may help accelerate convergence to optimal parameter sets.

Having focused primarily on computational examples, it would be interesting to see how the algorithm fares in extracting particularly elusive parameters from noisy experimental data---especially in setups where trajectories are readily available. Examples may include levitated nanoparticles, where one could measure nonlinearities of the confining potential, or protein-folding dynamics, to gauge the quality of reaction coordinates and extracted free energy landscapes. 
Such applications would greatly expand the preexisting catalogue of FT-based analysis tools and high-precision measurement techniques \cite{west2006free, hummer2010free, liphardt2002equilibrium, hayashi2010fluctuation, camunas2017experimental}.
Moreover, the white-noise DFT may also aid in assessing the relevance of selected parameters and variables on different timescales. For instance, the frequently employed overdamped representation of particle dynamics is bound to break down for sufficiently small time steps. The DFT can help in estimating the timescale below which the neglected degrees of freedom become relevant again to the description of the particle's propagation. Such a procedure could also help in the selection of adequate integration time steps for numerical applications. 

Finally, it should be noted that white noise itself breaks down as an approximation at sufficiently small timescales. This circumstance may motivate the theorem's generalization to other forms of noise, be it white noise with non-Gaussian statistics or stochastic forces with finite memory---including generalized Langevin equations with non-trivial memory kernels \cite{mori1965transport,kubo1966fluctuation}. Furthermore, in its current form the noise-symmetry can only be applied to classical trajectories. It would be interesting to see whether comparable laws could be unearthed for noisy quantum systems, similar to the extension of time-reversal FTs to the quantum regime \cite{talkner2007tasaki, campisi2009fluctuation}.

\begin{acknowledgments}
This research was supported by the Austrian Science Fund (grant no.\ I3163-N36) and by the Swiss National Science Foundation (SNF) through the NCCR-QSIT programme (grant no. 51NF40-160591).
\end{acknowledgments}

\section*{Data Availability}

The relevant datasets generated and analyzed throughout this work are available from the corresponding author upon reasonable request.

\appendix

\section{Beyond the It\^{o} Convention}

In Sec.~\ref{sec:DFT} we have derived the white-noise DFT under the assumption of the It\^{o} prescription. In this section, we seek to extend the generality and applicability of our results to arbitrary integration conventions. 
Integration conventions are specified by a parameter $\alpha \in [0,1]$ that uniquely prescribes the discretization of Eq.~\eqref{eq:langevin} \cite{lau2007state,risken1989fpe}. The value of $\alpha$ determines where the functions $f_i$ and $g_{ij}$ are evaluated to compute the evolution of $q_i$ throughout a small time step $\Delta t$, imposing
\begin{equation}
	\Delta q_i = f_i(\mathbf{q} +\alpha\Delta \mathbf{q})\Delta t + \sum_ j g_{ij}(\mathbf{q} +\alpha\Delta \mathbf{q}) \Delta W_{j}.
	\label{eq:ito}
\end{equation}
Physical trajectories typically follow the Stratonovich convention $\alpha=1/2$ rather than the It\^{o} prescription $\alpha = 0$ as a consequence of the Wong-Zakai theorem \cite{wong1965convergence}. Fortunately, it is possible to map Langevin equations from one integration convention to another \cite{gardiner2009stochastic, lau2007state}:
a system of Langevin equations as discretized in Eq.~\eqref{eq:ito} for arbitrary $\alpha$
is statistically equivalent with the system of (discretized) It\^{o}-SDEs
\begin{eqnarray}
	\Delta q_i &=& \left[ f_i(\mathbf{q})  + \alpha\sum_{j,k}  g_{jk}(\mathbf{q}) \frac{\partial g_{ik}(\mathbf{q})}{\partial q_j} \right]\Delta t  \nonumber\\
	&&+ \sum_ j g_{ij}(\mathbf{q}) \Delta W_{j}.
	\label{eq:itoTrafo}
\end{eqnarray} 
We point to Ref.~\cite{gardiner2009stochastic} for a detailed derivation of Eq.~\eqref{eq:itoTrafo}, as it lies beyond the scope of this work. The equation demonstrates that the conversion between different $\alpha$-prescriptions simply entails an appropriate modification of the deterministic term, sometimes called the thermodynamic drift. It is proportional to $\alpha$, or more generally $\alpha$-difference, and vanishes in the case of additive noise due to the derivatives in its expression. After modifying the drift term $f_i$ associated with convention $\alpha$ according to Eq.~\eqref{eq:itoTrafo}, one can once again apply the white-noise DFT as delineated in the main text.

Alternatively, it is also possible to formulate DFTs for arbitrary $\alpha$ by employing the respective expressions of the transition probability \cite{lau2007state}. Although doing so poses no additional conceptual difficulty, the expressions involved lose some simplicity and offer seemingly no noteworthy advantage over the It\^{o}-based treatment.

\section{Simulation Methods}

The simulations conducted throughout the course of the present study can be broadly split into three steps: trajectory generation, preparation of $Z_i$ statistics, and subsequent extraction of the $\Sigma(Z_i)$ symmetry function. The last step is performed either as consistency check or to extract the slope $R$ for the purpose of parameter optimization. In this section, we will briefly outline the methods we have employed to perform these tasks. 

As mentioned in Sec.~\ref{sec:trajCorrelations}, we can either generate an ensemble of trajectory segments or a single long trajectory to compute representative samples of the marginal displacement distributions $P(\Delta Z_i,w_i)$. In particular, the non-conservative system of Sec.~\ref{Sec:Overdamped_Systems} relies on the trajectory ensemble approach, resetting the particle once it leaves a selected perimeter around the origin. Here, all trajectory segments begin at the origin with a velocity drawn from the Maxwell-Boltzmann distribution, prescribing $P(\mathbf{q}_0)$. In contrast, the two systems of Sec.~\ref{Sec:Underdamped_Systems} follow the single trajectory approach. The initial state can be chosen arbitrarily in this case. All trajectories generated in this work use the symplectic OVRVO-integrator devised by Sivak et al.~\cite{sivak2014time}.

We proceed to compute an array of weighted displacements $\Delta Z_i$ from the trajectory data, the precise expressions for each test system are given in the respective subsections. At this point subsequent realizations of $\Delta Z_i$ are still correlated by construction; the array is thus randomly shuffled before summing several $\Delta Z_i$ into a single realization of $Z_i$. For the extraction of $\Sigma(Z_i)$ it is numerically convenient to choose a number of addends $n$ that puts the average $\left \langle Z_i \right \rangle$ close to unity. The independent selection of addends from a sufficiently large sample implies the relation $\left \langle Z_i \right \rangle = n \left \langle \Delta Z_i \right \rangle$, making the selection of $n$ straightforward. If desired, reducing the time step $\Delta t$ will result in an approximately inversely proportional increase of $n$ as determined from this criterion. In the presented applications, we have merely ensured that the average of $Z_i$ resides in an appropriate order of magnitude. Next, we compute the corresponding $Z_i$-histograms. The bins should be constructed symmetrically around the origin $Z_i=0$. This alignment is necessary for the computation of the logarithmic probability ratio $\Sigma(Z_i)$. Finally, for parameter inference applications a linear curve of the form $RZ_i$ is fitted to $\Sigma(Z_i)$ to extract the slope $R$ as discussed in Sec.~\ref{Sec:Inference}. We have used a standard least-squares method to this end.

\bibliographystyle{naturemag}
\bibliography{referencesShort}

\end{document}